\documentclass[twocolumn,amsmath,nofootinbib,amssymb]{revtex4}
\usepackage{float}
\usepackage{epsfig}
\usepackage{graphicx}
\usepackage{dcolumn}
\usepackage{bm}
\usepackage{color}
\usepackage{tabularx}
\usepackage{cleveref}

\def\thebiblio#1{
\begin{center}\bf \large References
\end{center}
\list
{[\arabic{enumi}]}{\settowidth\labelwidth{#1.}\leftmargin\labelwidth
 \advance\leftmargin\labelsep
 \usecounter{enumi}}
 \def\newblock{\hskip .11em plus .33em minus -.07em}
 \sloppy
 \sfcode`\.=1000\relax}

\begin{document}
\preprint{}
\title{%
Isotropy theorem for cosmological Yang-Mills theories
}

\author{J. A. R. Cembranos, A.\,L.\,Maroto and S. J. N\'u\~nez Jare\~no}
\address{Departamento de  F\'{\i}sica Te\'orica I, Universidad Complutense de Madrid, E-28040 Madrid, Spain}

\date{\today}

\begin{abstract}
We consider homogeneous non-abelian vector fields with general potential terms in an
expanding universe. We find a mechanical analogy with a system of $N$
 interacting particles (with $N$ the dimension of the gauge group) moving in three dimensions under the action of a central potential. In the case of bounded and rapid evolution compared to the rate of expansion, we show by making use of a generalization of the  virial theorem that for arbitrary potential and polarization pattern, the average energy-momentum tensor is always diagonal and isotropic despite the intrinsic anisotropic evolution of the vector field. We consider also the case in which a
gauge-fixing term is introduced in the action and show that the average equation
of state does not depend on such a term.  Finally, we extend the results to arbitrary background
geometries and show that  the average energy-momentum tensor of a rapidly evolving
Yang-Mills fields  is always isotropic and has the perfect fluid form for any locally inertial observer.
\end{abstract}

\maketitle

\section{Introduction}

The possibility that some of the unknown components of the universe
at different epochs (inflaton field, dark matter or dark energy) could
be described by means of homogeneous vector fields rather than scalar fields
have conflicted traditionally with the stringent limits on isotropy imposed by CMB observations
\cite{Ford,Armendariz,vectorinflation,Peloso,DE,VT,Dimopoulos,Nelson,Golovnev,Gflation,Soda}. The only scenarios
in which these limitations could be evaded  are those in which either  only the temporal components of the vector
fields
are present
or,  in the case in which spatial components are also
evolving, if the vector configuration guarantees an isotropic energy-momentum tensor.  Two possible
isotropic configurations discussed in the literature  are the presence of a triad of mutually orthogonal vectors \cite{solution, Galtsov, Zhang}
or the existence of a large $N$ number of randomly oriented
vectors, so that the average isotropy violation is kept relatively small of order $1/\sqrt{N}$  \cite{Golovnev}.

However, recently  \cite{Isotropy} a new possibility was devised in which a general isotropy theorem for vector fields
was proved. Provided vector field evolution is bounded  and rapid  compared to the rate of expansion,
it was shown that the average energy-momentum tensor is always diagonal and isotropic for any kind
of initial configuration. A typical example of rapid bounded evolution corresponds to  massive
vector fields with  masses larger than the Hubble parameter.
The theorem was proved in the case of abelian fields with standard Maxwell kinetic term.
In this work we are interested
in extending it to more general non-abelian gauge theories in which gauge-fixing terms could
also be present.

Unlike abelian theories, the  presence of multiple vector fields in Yang-Mills theories allows to implement in a natural
way the triad configuration mentioned before. Thus, in  \cite{Galtsov, Zhang} it was found that
for SU(2) groups, homogeneous and isotropic configurations compatible with FRW backgrounds
can be parametrized with a single function of time.   Different proposals for dark energy or
inflationary models based on
this triad solution with different types of actions have been considered in the literature \cite{Galtsov,Zhang,Armendariz,Chromoinflation}.
However apart from this particular type of solution no further isotropic configuration has been
proposed to date.  The theorem presented in this work
applies quite generally without resorting to particular solutions by means of a generalization of
the classical virial theorem and allows to ensure homogeneity and isotropy in average irrespective
of the concrete dynamics of the non-abelian fields for arbitrary semi-simple groups.

The paper is organized as follows, in Section II we introduce Yang-Mills theories in FRW backgrounds and
obtain the corresponding equations of motion and energy-momentum tensor components. In Section III
we introduce the generalized version of the virial theorem and show that  in average the energy-momentum tensor
 is
diagonal and isotropic for semi-simple groups. As an example we compute the
average equation of state for certain SU(2) solutions. In Section IV we extend these results to
actions including gauge-fixing terms and obtain the average equation of state in
different cases. We also consider the case with temporal components  only.
Section V is devoted to the extension to more general background metrics and
finally Section VI contains the main conclusions of the work.

\section{Yang-Mills theories}
Starting from a theory whose action has a global symmetry, if we would like to make it invariant under local symmetry transformations, we must add new bosons and define a covariant derivative:
\begin{equation}
D_{\mu}=\partial_{\mu}+A_{\mu}\;;\; A_{\mu}=-i g A^a_{\mu}T^a\;,
\end{equation}
where $T^a \in \mathcal{G}$ , $a=1\dots N$ are the symmetry group generators.  We will assume a compact semi-simple Lie group with a
finite number of generators. In this case, it is always possible to find an orthonormal basis  for the group generators for which:
\begin{equation}
 \mbox{Tr}(T^a T^b)=\frac{1}{2}\delta^{a b}\;.
\end{equation}
and the structure constants defined as:

\begin{equation}
\left[ T^a , T^b \right] = i c_{a b c} T^c\;;
\end{equation}

are  totally antisymmetric (see \cite{Ticciati}). This property will be very helpful for our future discussion.

In order to write a kinetic term for gauge bosons, we consider the curvature tensor
associated to the covariant derivative:
\begin{equation}
F_{\mu \nu}\equiv \left[ D_{\mu},D_{\nu} \right]=\partial_{\mu} A_{\nu}- \partial_{\nu} A_{\mu}+ \left[ A_{\mu},A_{\nu} \right]\;.
\end{equation}
Notice that this expression is still valid in curved space-time due to the antisymmetry properties of the
$F_{\mu\nu}$ tensor. The corresponding components read:
\begin{eqnarray}
F_{\mu \nu} &\equiv &- i g F^a_{\mu \nu} T^a\nonumber \\
 F^a_{\mu \nu}&=& \partial_{\mu} A^a_{\nu}-\partial_{\nu}A^a_{\mu} + g c_{a b c} A^b_{\mu} A^c_{\nu}\;.
\end{eqnarray}


Finally, the Yang-Mills lagrangian density reads:
 \begin{eqnarray}
 \mathcal{L}_{\text{kinetic}}= \frac{1}{2 g^2} \mbox{Tr} \left( F_{\mu \nu} F^{\mu \nu}  \right ) = - \frac{1}{4} F^a_{\mu \nu} F^{a \ \mu \nu}\;.
\label{kinetic}
 \end{eqnarray}

In this work we will study Yang-Mills theories with a potential of the form $V (M_{a b} A_{\rho}^a A^{b \rho})$,
with $M_{ab}$ a constant symmetric matrix, so  that  gauge symmetry is explicitly broken. Thus, the action in general
curved space-times will read:
\begin{equation}
\mathcal{S} = \int d^4x \sqrt{g}\left(- \frac{1}{4} F^a_{\mu \nu} F^{a \ \mu \nu}- V (M_{a b} A_{\rho}^a A^{b \rho})\right)\;.
\end{equation}
The corresponding equations of motion are given by:
\begin{equation}
F^{a\; ; \nu}_{\mu \nu}-g c_{a b c}F^{b}_{\mu \nu} A^{c\; \nu} + 2V' M_{a b} A^{b}_{\mu} = 0\;,
\end{equation}
where \mbox{$V' = \frac{d V(x)}{d x}$}.

We will be interested in cosmological solutions for which the gauge fields will depend only on time $A_\mu^a(\eta)$
and the metric tensor will be given by the that of a
flat FLRW metric:
\begin{eqnarray}
ds^2 =a^2(\eta) \left( d\eta^2-  d\vec{x}^2 \right)\;,
\end{eqnarray}
where $\eta$ denotes the conformal time coordinate.
 The field equations expressed in components can be written as
\begin{eqnarray}
g c_{a b c}\dot{A}^b_i A^{c}_i+g^2 c_{a b c} c_{b d e} A^d_0 A^e_i A^c_i &+& 2 V' M_{a b} a^2(\eta) A^b_0=0, \nonumber \\ \label{mu0}
\end{eqnarray}
for $\mu=0$, and
\begin{eqnarray}
\ddot{A}^a_i&-&g c_{a b c}\left( 2 \dot{A}^b_{i}A^c_0 + A^b_i\dot{A}^c_0 \right )+ g^2 c_{a b c}c_{b d e} \left ( A^d_i A^e_0A^c_0 \right. \nonumber
\\
&-& \left. A^d_i A^e_j A^c_j \right ) - 2 V' M_{a b} a^2(\eta)A^b_i=0\;, \label{mui}
\end{eqnarray}
for $\mu=i$. Notice that there is no second time derivative of the temporal component in these equations, as
expected for the standard kinetic term we are using.
On the other hand,
the energy-momentum tensor is given by:
\begin{equation}
T^{\mu}_{\;\;\nu} = \left( \frac{1}{4} F^a_{\alpha \beta} F^{a \ \alpha \beta} + V \right)\delta^{\mu}_{\;\;\nu} - F^{a \, \mu \alpha} F^a_{\nu \alpha} - 2 V' M_{a b} A^{a \mu} A^b_{\nu}.
\end{equation}
Expressed in components,
\begin{eqnarray}
\rho &=& \frac{1}{2a^4(\eta)}\left( \dot{A}^a_i \dot{A}^a_i+2g c_{a b c} \dot{A}^a_i A^b_0 A^c_i
\right.\nonumber \\
&+& \left. g^2 c_{a b c} c_{a d e} A^b_0 A^d_0A^c_jA^e_j \right)
+\frac{g^2 c_{a b c}c_{a d e}}{4a^4(\eta)}A^b_i A^d_i A^c_j A^e_j\nonumber \\
&+&V-2V'M_{a b}\frac{A^a_0 A^b_0}{a^2(\eta)}\;,
\end{eqnarray}
\begin{eqnarray}
p_{k} &=& -T^k_k= \frac{1}{2a^4(\eta)}\left( \dot{A}^a_i \dot{A}^a_i
+2g c_{a b c} \dot{A}^a_i A^b_0 A^c_i \right. \nonumber \\
&+& \left. g^2 c_{a b c} c_{a d e} \left(A^b_0 A^d_0
- \frac{A^b_i A^d_i}{2} \right)A^c_jA^e_j \right)\nonumber \\
&-& \frac{1}{a^4(\eta)}\left( \dot{A}^a_k \dot{A}^a_k +2g c_{abc} \dot{A}^a_k A^c_k A^b_0 \right.\nonumber
 \\
&+& \left. g^2 c_{a b c}c_{a d e} \left( A^b_0A^d_0- A^b_j A^d_j  \right) A^c_k A^e_k \right)-V \nonumber
\\
&-& 2 V' M_{a b} \frac{A^a_k A^a_k}{a^2(\eta)} \;, \;\;k =\; 1,\; 2,\; 3\;;
\end{eqnarray}
\begin{eqnarray}
T^i_j &=& \frac{1}{a^4(\eta)}\left [ \dot{A}^a_i \dot{A}^a_j - g c_{a b c} \left ( \dot{A}^a_i A^b_j + \dot{A}^a_j A^b_i \right ) A^c_0    \right. \nonumber
\\
 &+& g^2c_{a b c} c_{a d e} A^b_i  A^d_j A^c_0 A^e_0 - g^2 c_{a b c}c_{a d e}A^b_i A^c_k A^d_j A^e_k \nonumber
\\
&+& 2 \left. V' M_{a b} a^2(\eta) A^a_i A^b_j \right ]\;, \;\;\; i\neq j \label{Tijgau}
\end{eqnarray}
\begin{eqnarray}
T^0_i &=& - \frac{A^a_i}{a^4(\eta)}\left (g c_{a b c}\dot{A}^b_j A^{c}_j+g^2 c_{a b c} c_{b d e} A^d_0 A^e_j A^c_j \right. \nonumber
\\
&+& \left. 2 V' M_{a b} a^2(\eta) A^b_0 \right)\;.
\end{eqnarray}
The average of the pressure along the three spatial directions will be denoted by:
\begin{eqnarray}
p \equiv \frac{1}{3} \sum_k p_k \;.\label{p}
\end{eqnarray}
 It is possible to show that, using the equations of motion of the temporal components (\ref{mu0}), the energy fluxes vanish, indeed:
\begin{eqnarray}
T^0_i &=& 0\;.
\end{eqnarray}

\section{Generalized virial theorem for non-abelian fields}
We will follow the method presented in
 \cite{Isotropy}, in order  to compute the  average  components of the energy-momentum tensor.
With that purpose, we will generalize the virial theorem in order to apply it to non-abelian fields, thus we define:
\begin{eqnarray}
G^{a b}_{ij}=\frac{\dot A^a_i A^b_j}{a^4(\eta)}, \;\;\; i,j=1,2,3; \;\; a,b=1\dots N\;.
\label{G}
\end{eqnarray}
Now \mbox{$A^a_i$} is alike the "i" position coordinate of the classical point particle "a". Thus, in the case $A^a_0 =0$, the problem is analogue to a mechanical system of $N$ interacting particles
in three dimensions, with $N$ the group dimension. Notice however that the isotropy theorem
we will show below is valid also for $A^a_0 \neq 0$.
\\
\\
Assuming a rapid evolution of the vector field, we can neglect the expansion of the universe at the field equations, the time derivative of the previous expression becomes:
\begin{eqnarray}
\dot G^{a b}_{ij}=\frac{\ddot A^a_i A^b_j}{a^4(\eta)}+\frac{\dot A^a_i \dot A^b_j}{a^4(\eta)},\;\;
 i,j&=&1,2,3; \nonumber \\a,b&=&1\dots N\;.
\end{eqnarray}
Integrating the last expression on $[0,T]$ where $T$ is larger than the typical time scale of
the vector field evolution $\omega^{-1}$ but smaller than  the typical time scale of the universe expansion $H^{-1}$, i.e.
 $H^{-1} \gg T \gg \omega^{-1}$, we obtain:
\begin{eqnarray}
\frac{G^{a b}_{ij}(T)-G^{a b}_{ij}(0)}{T}&=& \left \langle \frac{\ddot A^a_i A^b_j}{a^4(\eta)}+\frac{\dot A^a_i \dot A^b_j}{a^4(\eta)} \right \rangle \;. \label{interval}
\end{eqnarray}
If the motion is periodic or it is bounded, the left hand side of the equation vanishes and therefore:
\begin{eqnarray}
 \left \langle \frac{\ddot A^a_i A^b_j}{a^4(\eta)}+\frac{\dot A^a_i \dot A^b_j}{a^4(\eta)} \right \rangle =0 \;.
 \label{vircero}
\end{eqnarray}
 Using these equations, we will show in the following  that the average energy-momentum tensor is diagonal and isotropic.

From the off-diagonal part of the tensor (\ref{Tijgau}), noting that:
\begin{eqnarray}
\ddot{A}^a_{(i}A^a_{j)} &=& 2 g c_{a b c}\dot{A}^b_{(i} A^a_{j)}A^c_0 - g^2 c_{a b c}c_{b d e} \left ( A^d_{(i}A^a_{j)} A^e_0 A^c_0 \right. \nonumber
\\
&-& \left. A^d_{(i} A^a_{j)} A^e_k A^c_k\right ) + 2V' M_{a b} a^2(\eta) A^b_{(i} A^a_{j)}\; ,\label{gauvir}
\end{eqnarray}
where the parenthesis in the sub-index means symmetrization,  we can write:
\begin{eqnarray}
T^i_j&=&\frac{1}{a^4(\eta)} \left (\dot{A}^a_i \dot{A}^a_j + \ddot{A}^a_{(i} \dot{A}^a_{j)} \right)
\nonumber \\
&=&\frac{1}{a^4(\eta)} \left (\dot{A}^a_{(i} \dot{A}^a_{j)} + \ddot{A}^a_{(i} \dot{A}^a_{j)} \right),
\nonumber
\\
\mbox{with} \;\;
 i&\neq &j\;.
\end{eqnarray}
Using the average expression (\ref{vircero}), we directly obtain:
\begin{equation}
\langle T^i_j \rangle=\frac{1}{a^4(\eta)} \left ( \left \langle \dot{A}^a_{(i} \dot{A}^a_{j)} + \ddot{A}^a_{(i} \dot{A}^a_{j)}\right \rangle \right)= 0\;, \;\; i\neq j \;.
\end{equation}
We can prove the isotropy of the diagonal elements in a similar way. Indeed,  the diagonal spatial components of the
energy-momentum tensor are
\begin{eqnarray}
T^k_k = \left ( \frac{1}{4}F^a_{\rho \lambda} F^{a \; \rho \lambda} + V \right )- t^k_k, \;\;\;
k=1,2,3\,;
\end{eqnarray}
where $t^k_k$ is the anisotropic part of these components,
\begin{eqnarray}
t^k_k &=& \frac{1}{a^4(\eta)} \left [ \dot{A}^a_k \dot{A}^a_k - 2g c_{a b c} \dot{A}^a_k A^b_kA^c_0 \right.  \nonumber
\\
&+& g^2 c_{a b c}c_{a d e} \left ( A^b_k A^d_k A^e_0 A^c_0
-  A^b_k A^d_k A^c_j A^e_j \right) \nonumber \\
&+&\left. 2V' a^2(\eta) M_{a b} A^a_k A^b_k \right ], \;\;\;
k=1,2,3\,;
\end{eqnarray}
which is equal to the expression $T^i_j$ when $i = j$. Therefore
\begin{eqnarray}
t^k_k = \frac{1}{a^4(\eta)} \left [  \dot{A}^a_{(k} \dot{A}^a_{k)}+ \ddot{A}^a_{(k} A^a_{k)}\right ]\Rightarrow \langle t^k_k \rangle = 0, \;\;\; k=1,2,3\,;\label{tk}\nonumber \\
\end{eqnarray}
where again we have made use of (\ref{vircero}). Thus, we can conclude that the virial theorem guarantees the isotropy of the energy-momentum tensor and its diagonal form in average.

\subsection{Equation of state for SU(2) particular solutions}

For SU(2) groups some particular solutions  are already known in flat space-time, see \cite{solution}
where a triad solution was proposed:
\begin{eqnarray}
A^a_{i}&=& k^a_{i} f(\eta)\;;\;\text{with} \; \vec{k}^a\cdot\vec{k}^b = \delta^{a b}\;. \label{partsol}
\\ A^a_0&=&0\;.
\end{eqnarray}
 Notice that in those solutions the temporal components of the vector fields vanish. For abelian field theories the
vanishing of the temporal component comes naturally from the equation of motion, as it can be seen in (\ref{mu0}) by making $g=0$. In the non-abelian case it would be possible to have solutions with non-vanishing
time components, however in the present example we are not  considering this possibility. Notice also that this kind of solutions only depend on a single function of time. We have checked that for the above
ansatz, solutions also exists in an expanding background.

Let us define:
 \begin{equation}
 \tilde{V}= \frac{g^2}{4 a^4(\eta)} c_{a b c} c_{a d e} ( A^b_i A^d_i)(A^c_j A^e_j)+ V\left(-M_{a b}\frac{A^a_i A^b_i}{a^2(\eta)}\right)\;.
 \end{equation}
The average energy-momentum tensor components then take the form:
\begin{eqnarray}
\langle \rho \rangle &=&\left\langle \frac{1}{2} \frac{\dot{A}_i^a\dot{A}_i^a}{a^4(\eta)}+ \tilde{V} \right\rangle\;,
\\
\langle p_{k} \rangle &=& \langle p  \rangle =  \left\langle \frac{1}{2} \frac{\dot{A}_i^a\dot{A}_i^a}{a^4(\eta)} - \tilde{V} \right\rangle\;,
\\
T^0_i &=& 0 \;,\label{T0i}
\\
T^i_j &=& 0\;,\;\;\; i\neq j\;.
\end{eqnarray}
Notice that this particular example corresponds to the triad case
discussed before for which the energy-momentum tensor is manifestly isotropic.
 For simplicity we will assume a power law potential:
 \begin{eqnarray}
 V\left ( M_{a b} A^{a}_{\rho} A^{b\;\rho} \right)=\frac{1}{2}(-M^2 A^a_{\rho} A^{a \rho})^n\;. \label{powlaw}
 \end{eqnarray}
For the particular ansatz  (\ref{partsol}), we get:
\begin{equation}
\langle \rho \rangle = \left\langle\frac{3}{2} \frac{\dot{f}^2}{a^4(\eta)} + \frac{(3 M^2 f^2)^n}{2 a^{2n}(\eta)} + \frac{3}{2}\frac{g^2}{a^4(\eta)} f^4 \right\rangle,
\end{equation}
\begin{equation}
 \langle p\rangle = \left\langle \frac{3}{2} \frac{\dot{f}^2}{a^4(\eta)} -  \frac{(3 M^2 f^2)^n}{2 a^{2n}(\eta)} - \frac{3}{2} \frac{g^2}{a^4(\eta)} f^4 \right\rangle,
\end{equation}
\begin{eqnarray}
T^0_i &=& 0,
\\
T^i_j &=& 0\;,\;\;\; i\neq j\;.
\end{eqnarray}
We can write:
\begin{equation}
\left \langle \frac{3 \dot{f}^2}{a^4(\eta)}\right \rangle = \langle \rho + p \rangle =(\gamma+\gamma_p) \langle \rho \rangle,
\end{equation}
where $\gamma$ is the average value over an oscillation and $\gamma_p$ the periodic part of the oscillation, that can be neglected (see \cite{Turner}). Calling $r = \frac{f}{a}$ and  taking the scale factor as a constant $\dot{r}=\frac{\dot{f}}{a}$,
we can define an effective potential:
\begin{eqnarray}
V_{eff}(r)= \frac{1}{2} (3M^2 r^2)^{n}  + \frac{3}{2} g^2r^4\,,
\label{veff}
\end{eqnarray}
such that:
\begin{eqnarray}
\rho=\frac{3}{2} \frac{\dot r^2}{a^2(\eta)}+V_{eff}(r)\;.
\label{rhoveff}
\end{eqnarray}
Notice that the non-abelian contributions appear as a quartic term in the effective potential,
so that in the strong coupling limit, we expect the equation of state to correspond to that of
a scalar field with a quartic potential, i.e. that of radiation  \cite{Turner}.

\begin{figure}[t]
  \includegraphics[width=70mm,angle=0]{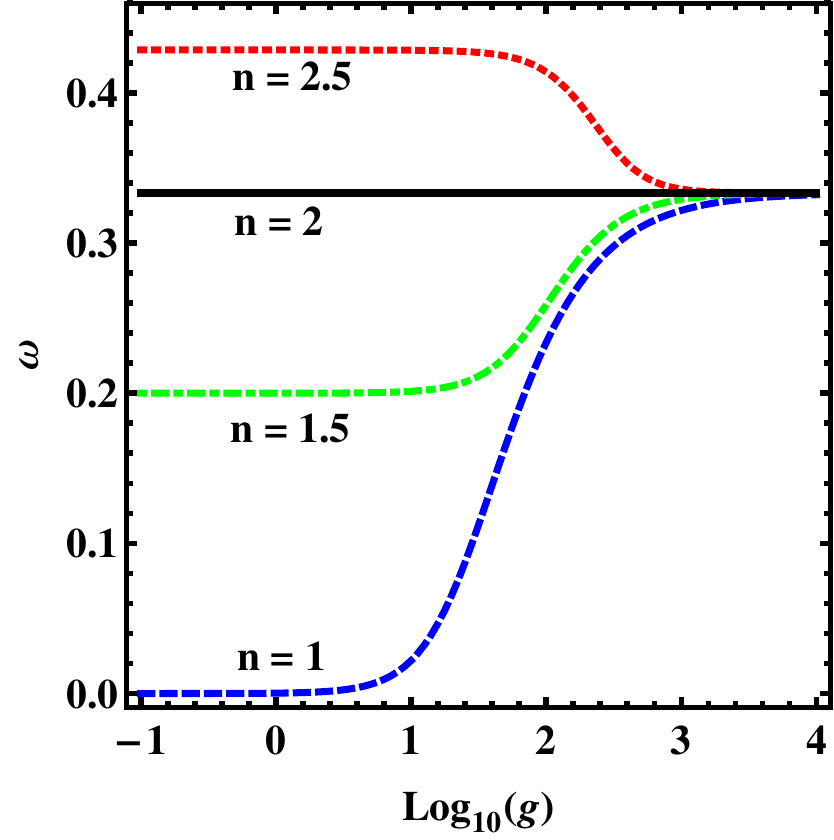}
 \caption{In this figure, it is shown how the equation of state $\langle p \rangle = \omega \langle \rho \rangle$ changes  by varying $g$ in the SU(2) non-abelian case with potential (\ref{powlaw}), for a given  energy density and $M$ parameter. The different lines correspond to different values of the potential exponent $n$ in (\ref{veff}). We can observe that for large $g$ values, the non-abelian term dominates and the equation of state approaches the radiation behaviour independently of $n$. On the contrary, for small $g$, the abelian case is recovered \cite{Isotropy}.}\label{Figg}
 \end{figure}

In order to compute the equation of state, we first obtain
 $\gamma$  following \cite{Turner} as:
 \begin{equation}
 \gamma =\frac{2}{\langle \rho \rangle}
\frac{\int^{r_+}_{r_-}  dr \sqrt{ \rho - V_{eff}(r)}}{\int^{r_+}_{r_-}\frac{dr}{\sqrt{\rho - V_{eff}(r)}}},
 \end{equation}
 where $r_+$ and $r_-$  are the turning points of the effective potential where $\dot r=0$.
The corresponding equation of state is $\langle p \rangle = w \langle \rho \rangle$, being $w= \gamma -1$. The integrals cannot be computed analytically but the numerical results\footnote{Now, unlike the abelian case, $ f\, d V_{eff}/df $ is not proportional to $V_{eff}$, except when $n=2$, and that is the reason why the virial theorem can not be used to obtain
the equation of state analytically as done in \cite{Isotropy}} show that $w$ depends on $n$, $\rho$, $g$ and $M$ varying between radiation for large $g$ (the new non-abelian term dominates in (\ref{veff})) and the corresponding $\omega$ of the abelian case \cite{Isotropy} for a given $n$ when $g$ is small (see Fig. \ref{Figg}).
 \begin{figure}[t]
 \includegraphics[width=70mm,angle=0]{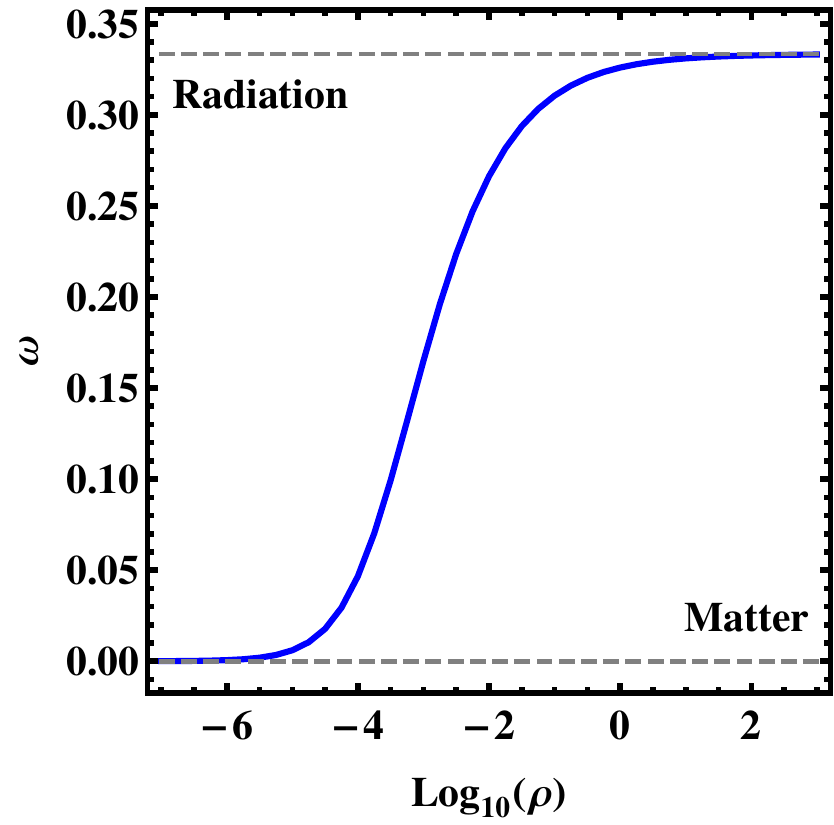}
 \caption{In this figure, it is shown how the equation of state $\langle p \rangle = \omega \langle \rho \rangle$ changes  by varying $\rho \left[ M^4 \right]$ in the SU(2) non-abelian case, for $g = 50$, $n=1$ and $M=1$. It can be seen how the behaviour of the higher power dominates at high energy density while the contrary happens when the energy density is low.} \label{Figrho}
 \end{figure}
\\
\\
The behaviour of the equation of state depends also on the energy density. If the energy density decreases as the
universe expands so that the amplitude of the field oscillations decreases in time, then at late times the smallest power
in the effective potential (\ref{veff}) will dominate. Thus, for power-law potentials we can have two possibilities: for $n \geq 2$, the potential
term will contribute at high energies (early times) and $\omega=\frac{n-1}{n+1}$, whereas at low energies (late times),
the non-abelian quartic term will dominate and $\omega=\frac{1}{3}$. For  $n < 2$, the situation is reversed and
at high energies, we have domination of the non-abelian term and  $\omega=\frac{1}{3}$, whereas at
low energies  $\omega=\frac{n-1}{n+1}$ (see Fig. \ref{Figrho} and reference \cite{Isotropy}).

\section{Yang-Mills theory with a gauge-fixing term}
As is well-known the standard kinetic term for gauge fields (\ref{kinetic}) does not contain a time
derivative of the temporal components which implies that no conjugate momenta can be defined for those fields.
Accordingly, temporal components cannot be quantized along with the spatial ones, thus preventing  an explicitly covariant
quantization of the theory.  Covariant quantization thus requires the modification of the kinetic term by including
new gauge-breaking terms \cite{Itzykson}. Those terms can be included directly in the action as in the Gupta-Bleuler formalism
or they could appear as a consequence of the selection of a representative  within
each gauge orbit in the
path integral approach. Restricting to the case of quadratic   gauge-breaking terms the
action reads:
\begin{equation}
\mathcal{L} = - \frac{1}{4} F^a_{\mu \nu} F^{a \ \mu \nu}+ \frac{\xi}{2}(\nabla_{\rho} A^{a \;\rho})^2- V (M_{a b} A_{\rho}^a A^{b \rho})\;.
\end{equation}
We can express the covariant equations of motion as
\begin{equation}
F^{a\; ; \nu}_{\mu \nu}-g c_{a b c}F^{b}_{\mu \nu} A^{c\; \nu} + 2V' M_{a b} A^{b}_{\mu}  + \xi\left(  A^{a\;;\rho}_{\rho}\right)_{;\mu} = 0\;.
\end{equation}
By introducing the homogeneity condition, this expression is reduced to:
\begin{eqnarray}
\xi \ddot{A}^a_0 &+& g c_{a b c}\dot{A}^b_i A^{c}_i+g^2 c_{a b c} c_{b d e} A^d_0 A^e_i A^c_i
\label{eomgaut}\\
&+& \left(2 \xi \left(\frac{\ddot{a}}{a}  - 3 \frac{\dot{a}^2}{a^2} \right) \delta_{a b}+2 V' M_{a b} a^2(\eta)\right )A^b_0=0 \;, \nonumber
\\
\nonumber
\\
\ddot{A}^a_i&-&g c_{a b c}\left( 2 \dot{A}^b_{i}A^c_0 + A^b_i\dot{A}^c_0 \right )+ g^2 c_{a b c}c_{b d e} \left ( A^d_i A^e_0A^c_0 \right. \nonumber
\\
&-& \left. A^d_i A^e_j A^c_j \right ) - 2 V' M_{a b} a^2(\eta) A^b_i=0\;. \label{eomgaus}
\end{eqnarray}
In order to have oscillations we must restrict the parameter space and initial conditions. In the case of a power law potential like (\ref{powlaw}) with an odd power, those conditions  simply require $\xi < 0$. Notice that we do not need to impose any condition over the temporal component in order to ensure isotropy.
\\
The energy-momentum tensor has the form:
\begin{eqnarray}
T_{\mu \nu}&=& \left( \frac{1}{4} F^{a}_{\rho \lambda} F^{a \;\rho \lambda}+ V \right ) g_{\mu \nu} - F^{a\; \rho}_{\mu} F^{a}_{\nu \rho} \nonumber
\\
&-& 2 V' M_{a b} A^{a}_{\mu} A^{b}_{\nu} + \frac{\xi}{2}\left[ g_{\mu \nu} \left[ (\nabla_{\rho}A^{a \; \rho})^2 \right. \right.
\\
&+& \left. \left.  2 A^{a}_{\lambda} \nabla^{\lambda} (\nabla_{\rho}A^{a \; \rho}) \right]- 4 A^a_{(\mu }\nabla_{ \nu)} (\nabla_{\rho} A^{a \; \rho}) \right]\;,\nonumber
\end{eqnarray}
where the parenthesis in the sub-index means symmetrization. If $\xi<0$, the energy density associated to the temporal component of the field is generally  negative defined. This is consistent with the standard interpretation of $A^a_{0}$ as a ghost field. The presence of ghosts is a potential 
problem in vector field theories and has to be carefully studied case by case \cite{Peloso}.

In the following, we will show that the introduction of the gauge breaking term does not spoil the
averaged isotropy of the energy-momentum tensor.  Let us first consider the energy fluxes. We see that:
\begin{eqnarray}
T^0_i &=& - \frac{A^a_i}{a^4}\left [ \xi \ddot{A}^a_0+g c_{a b c}\dot{A}^b_j A^{c}_j+g^2 c_{a b c} c_{b d e} A^d_0 A^e_j A^c_j \right. \nonumber
\\
&+& \left. \left(2 \xi \left(\frac{\ddot{a}}{a}  - 3 \frac{\dot{a}^2}{a^2} \right) \delta_{a b}+2 V' M_{a b} a^2(\eta)\right )A^b_0 \right ]= 0\;,\nonumber \\
\end{eqnarray}
where in the last step we have used the equations of motion of the temporal components (\ref{eomgaut}).
On the other hand, the new $\xi$ term does not contribute either to $T^i_j$ or $t^k_k$. Therefore, following the same procedure of the previous section (see Eqs. (\ref{G}) to (\ref{tk})), we can conclude that the virial theorem also guarantees in this case the isotropy of the energy-momentum tensor and its diagonal form in average if the spatial fields oscillate fast enough compared with the expansion rate of the universe.
\subsection{Equation of state}
In order to obtain  the equation of state we will proceed in a different way as it has been done
in the previous section. We take the trace of the energy-momentum tensor:
\begin{eqnarray}
T^{\mu}_{\mu}&=& \rho - 3p = 4V - 2V'M_{a b}A^{a\; \mu} A^b_{\mu}\nonumber
\\
&+& 2 \xi \left[ (\nabla_{\rho}A^{a \; \rho})^2 + A^{a \mu}\nabla_{\mu}(\nabla_{\rho}A^{a \; \rho})\right]\;. \label{trace}
\end{eqnarray}
with $p$ the average pressure along the three spatial directions defined in (\ref{p}). Thus, we have:
\begin{eqnarray}
\rho - 3p  &=&  4V - 2V'M_{a b}A^{a\; \mu} A^b_{\mu} \nonumber
\\
&+& 2 \xi  \nabla^{\lambda} \left(A^{a}_{\lambda} \nabla_{\rho}A^{a \; \rho}\right) \;.
\end{eqnarray}
Notice that the fast oscillations condition is not mandatory for the temporal part since it
does not contribute to the anisotropies. Nevertheless, if the oscillations are indeed faster than the expansion rate of the universe, taking the average, we get:
\begin{eqnarray}
(1- 3 \omega)\langle \rho \rangle &=& \left\langle 4V - 2V'M_{a b}A^{a\; \mu} A^b_{\mu} \right\rangle \nonumber
\\
&+& 2 \xi \left \langle \nabla^{\lambda} \left({A^{a}_{\lambda} \nabla_{\rho}A^{a \; \rho}}\right) \right\rangle\;.
\end{eqnarray}
Notice  that the $\xi$ term, when neglecting the scale factor derivatives, is nothing but  the temporal derivative of the $G^{a a}_{0 0}$ function used in the virial theorem which in average vanishes, i.e.:
\begin{equation}
G_{0 0}^{a b}=\frac{\dot{A^a_{0}}A^b_{0}}{a^2(\eta)}\;,\;\;\; a,b=1\dots N,
\end{equation}
and as  in the spatial case (\ref{vircero}), we get:
\begin{eqnarray}
\langle \dot G_{0 0}^{a b}\rangle= \left \langle \frac{\ddot A^a_0 A^b_0}{a^4(\eta)}+\frac{\dot A^a_0 \dot A^b_0}{a^4(\eta)} \right \rangle =0\;,\;\;\; a,b=1\dots N.
\label{vir}
\end{eqnarray}

Thus, we can give an expression for $\omega$ depending only on the average of the potential and the energy density:
\begin{equation}
\omega = \frac{1}{3}+ \frac{\left\langle 2V'M_{a b}A^{a\; \mu} A^b_{\mu}-4V \right\rangle}{3 \langle \rho \rangle}\;.
\end{equation}
Considering a power law potential, the expression reduces to:
\begin{eqnarray}
\omega = \frac{1}{3}+\frac{2(n-2)}{3} \frac{\left\langle V \right\rangle}{ \langle \rho \rangle}\;.\label{omegatemp}
\end{eqnarray}
As it can be seen the solution for a potential with $n=2$ can be obtained analytically, it behaves as radiation $\omega = \frac{1}{3}$. That is the expected result considering the behaviour of the same case without gauge-fixing term, see Fig. \ref{Figg}.

\subsubsection{Abelian case}

In the abelian case (or $g \rightarrow 0$) we can obtain the exact expression for the average
equation of state by following the virial method despite $A^a_0 \neq 0$.
Using (\ref{vir}) and (\ref{vircero}) and the equations of motion (\ref{eomgaut}) and (\ref{eomgaus}), we have:
\begin{eqnarray}
\left \langle \dot{A}^a_0 \dot{A}^a_0 \right \rangle &=& \frac{2M_{a b} a^2(\eta)}{\xi}\left \langle V' A^a_0 A^b_0\right \rangle \;, \label{Vtemporal1}
\\
\left \langle \dot{A}^a_i \dot{A}^a_i \right \rangle &=& -2M_{a b} a^2(\eta) \left \langle V' A^a_i A^b_i \right \rangle \;,\label{Vtemporal2}
\end{eqnarray}
and the average energy density  can be written as
\begin{eqnarray}
\langle \rho \rangle &=& \left \langle \frac{1}{2} \frac{\dot{A}^a_i \dot{A}^a_i}{a^4(\eta)} + \frac{\xi}{2} \frac{\dot{A}^a_0 \dot{A}^a_0}{a^4(\eta)} + V \right. \nonumber
\\
 &-& \left. \frac{1}{a^4(\eta)} (\xi \ddot{A}^a_0+2V'M_{a b}a^2(\eta) A^b_0)A^a_0\right \rangle\;.
\end{eqnarray}
Using  (\ref{eomgaut})  and neglecting time
derivatives of the scale factor, we see that the last term vanishes and finally we get:
\begin{eqnarray}
\langle \rho \rangle = \left \langle \frac{1}{2} \frac{\dot{A}^a_i \dot{A}^a_i}{a^4(\eta)} + \frac{\xi}{2 } \frac{\dot{A}^a_0 \dot{A}^a_0}{a^4(\eta)} + V
\right \rangle\;.
\end{eqnarray}

Then using (\ref{Vtemporal1}) and (\ref{Vtemporal2}), we obtain:
\begin{equation}
\langle \rho \rangle = (n+1)\left \langle V \right \rangle\;.
\end{equation}
Introducing this expression in (\ref{omegatemp}), we reach the same result as \cite{Turner} for a scalar field and \cite{Isotropy} for an abelian theory without gauge-fixing term:
\begin{eqnarray}
\omega = \frac{n-1}{n+1}\;. \label{vireom}
\end{eqnarray}

\subsubsection{Non-abelian case}
 For non-abelian theories we must compute the average of the potential and the energy density in (\ref{omegatemp}).  As commented before, we cannot apply  the virial theorem as in the abelian case for simple power law potentials  because now
some non-abelian terms (those multiplied by $g^2$ ) act as additional potential terms. Thus the effective potential becomes a
sum of power laws, and $V_{eff}' A^2$ is not proportional to $V_{eff}$. In any case, if we assume $0<\left\langle V \right\rangle / \langle \rho \rangle<1$, we can constrain the value of $\omega$:
\begin{equation}
\left. \left.\begin{array}{c}
\frac{2n-3}{3}
\\
\\
\frac{1}{3}
\end{array} \right\} < \omega < \left \{  \begin{array}{c}
\frac{1}{3}\;\;\;\;\;\;,\; \text{for}\;n < 2\;;
\\
\\
\frac{2n-3}{3}\;,\; \text{for}\;n >2\;.
\end{array}
\right.\right.
\end{equation}

\subsection{Particular case: $A^a_{\mu}=A^a_{0}(\eta)\delta^{0}_{\mu}$}

Let us consider the case in which  the only relevant component of the homogeneous Yang-Mills theory with a gauge-fixing term ( i.e. $\xi \neq 0$ ) is the temporal part of the fields,
\begin{equation}
A^a_{\mu}=A^a_{0}(\eta)\delta^{0}_{\mu}\;.
\end{equation}
The equations of motion become:
\begin{equation}
\ddot{A}^a_0 +\left(2 \left(\frac{\ddot{a}}{a}  - 3 \frac{\dot{a}^2}{a^2} \right) \delta_{a b}+2 V' \frac{M_{a b}}{\xi} a^2(\eta)\right )A^b_0=0 \;,
\end{equation}
with  no contribution from the non-abelian terms. This equation is similar to the standard scalar field equation, although with variable mass depending on the scale factor derivatives.
However the non-zero components of the  energy-momentum tensor are completely different,
\begin{eqnarray}
\rho&=&V+\frac{\xi}{2} \left ( \nabla_{\rho} A^{\rho} \right )^2 \underbrace{-2 V' M_{a b} A^a_0 A^b_0-\xi A^{a \;0} \nabla_0 \left( \nabla_{\rho} A^{a \;\rho} \right)}_{=0\;,\;\text{by e.o.m.}} \nonumber
\\
 &=& V+\frac{\xi}{2} \left ( \nabla_{\rho} A^{\rho} \right )^2\;;
 \\
 \nonumber
 \\
 p &=& -V -\frac{\xi}{2} \left ( \nabla_{\rho} A^{\rho} \right )^2 -\xi A^{a \;0} \nabla_0 \left( \nabla_{\rho} A^{a \;\rho} \right)\;.
\end{eqnarray}
Adding them:
\begin{equation}
\rho + p =-\xi A^{a \;0} \nabla_0 \left( \nabla_{\rho} A^{a \;\rho} \right)= 2 V' M_{a b} A^{a\;0} A^b_0\;,
\end{equation}
where in the last step the equations of motion have been used. We can give an expression for the equation of state:
\begin{equation}
\omega = \frac{2 V' M_{a b} A^{a\;0} A^b_0}{\rho}-1\;.
\end{equation}
Now  the oscillatory behaviour is not required as the isotropy is guaranteed, so for a simple power-law potential, we get:
\begin{equation}
\omega = \frac{2 n}{ 1+\frac{\xi \left ( \nabla_{\rho} A^{\rho} \right )^2}{2 V} }-1\;.
\end{equation}
We see that contrary to the  scalar field standard behaviour, when the potential dominates $\omega=2n-1$; whereas $\omega=-1$ when the potential term is negligible.
When the field is  rapidly oscillating, we can use the virial equations (\ref{vireom})  and the average $\omega$ behaves as a standard scalar field $\omega =\frac{n-1}{n+1}$ since
\begin{equation}
\left \langle\frac{\xi \left ( \nabla_{\rho} A^{\rho} \right )^2}{V} \right \rangle = 2\, n\;,
\end{equation}
as it can be straightforwardly deduced from Eq. (\ref{Vtemporal1}).

\begin{figure*}[ht]
 \includegraphics[width=175mm,angle=0]{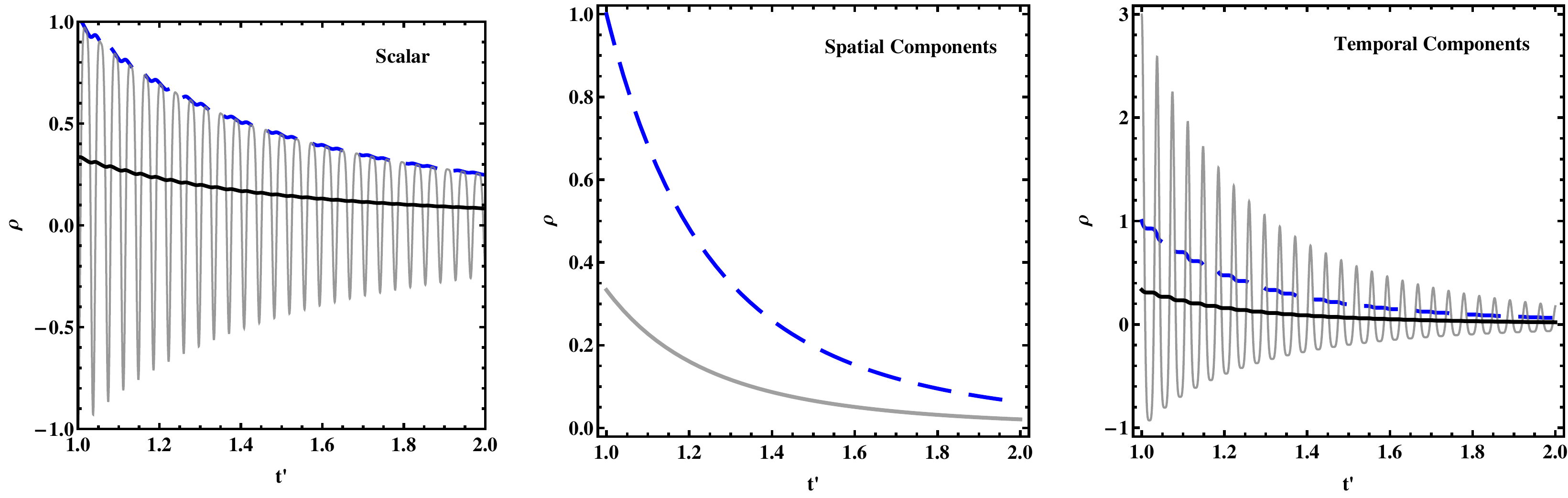}
 \caption{ In this figure the average pressure along the three spatial directions $p$ (grey line), energy density (blue dashed line) and average pressure $\langle p \rangle = \langle \rho \rangle/3$ (black line) of three different cases are shown. These cases are (from left to right) a scalar field with potential $V_{\phi}= M \phi^4$, a vector with spatial components only and a
 vector with temporal component only with
potential  $V_{A}= M \left (A^a_{\mu} A^{a \; \mu}\right)^2$, where $M=100$, $\xi = 1$, $g=0$. The y-axis is normalized to the initial value $\rho(t'_0)$ and time, $t'$, is in $H_0^{-1}$ units. We are considering a radiation dominated-universe. Notice that in the spatial components case
$\langle p \rangle = p$.}\label{Figpressure}
 \end{figure*}

This does not mean that the temporal component becomes a standard scalar field, but that in average its equation of state behaviour is the same. A good way to illustrate this is by studying the case $V_{A}= M \left (A^a_{\mu} A^{a \; \mu}\right)^2$ and its equivalent $V_{\phi}= M \phi^4$ for a scalar field. The conditions to have oscillations are $M>0$ and $\xi>0$. In this particular case, the pressure behaves completely different when the field is a scalar, a temporal component or a spatial component of
a  vector field; although, in average they behave in the same way.

In Figure \ref{Figpressure}, the different cases are shown for $g=0$ in a radiation-dominated universe. For a scalar field, the pressure will oscillate in the interval $- \rho < p < \rho$, so pressure is modulated by energy density. If we have a spatial component of a vector field ($A^a_0 =0$), the average pressure over the
three spatial directions
$p$ always equals $\rho/3$, and it does not oscillate. This odd behaviour can be explained by realising that the trace of the   energy-momentum tensor (\ref{trace}) vanishes for this potential if the temporal component is zero, then $\rho-3p=0$ must be satisfied at any time. Finally for a temporal component of a vector field ($A^a_i =0$), pressure oscillations exceed the energy density value on the contrary to the scalar case.

\section{Isotropy theorem for  Yang-Mills theories in a general background metric}
The previous results can be directly extended
to  general space-time geometries (not necessarily homogeneous) as was shown in \cite{Isotropy}. Thus,
let us consider a locally inertial observer at
$x_0^\mu=0$ and write
the metric tensor using Riemann normal coordinates around $x_0^\mu$ \cite{Petrov}:
\begin{eqnarray}
g_{\mu\nu}(x)=\eta_{\mu\nu}+\frac{1}{3}R_{\mu\alpha\nu\beta}x^\alpha x^\beta +\dots
\label{normal}
\end{eqnarray}
Let assume that the following conditions hold:
\begin{enumerate}
\item { The Lagrangian is restricted to the Yang-Mills form with or without a gauge-fixing term.}

\item {The vector field evolves rapidly:
\begin{eqnarray}
&&|R^\gamma_{\lambda\mu\nu}| \ll (\omega_{i}^{a})^{2} ,\;
\text{and} \;\;
|\partial_j A^a_{i}| \ll | \dot{A}^a_{i} | ,\;\;
\nonumber \\
&&\;\;\;\;\;\;\;\;\text{for}\;\; i,j=1,2,3\;;\;\text{and}\; a=1\dots N\;,
\end{eqnarray}
for  any component of the Riemann tensor.  $\omega^a_{i}$ is the characteristic frequency of $A^a_{i}$}

\item {$A^a_{i}$ and $\dot A^a_{i}$ remain bounded in the evolution.}
\end{enumerate}
The second condition implies that if we are only interested in time scales of
order $(\omega_i^{a})^{-1}$, then we are in a normal neighbourhood and we can neglect the second
term in (\ref{normal}) and also work with a homogeneous vector field.   In such a case,  it is possible to
rewrite all the above equations with $a(\eta)=1$. Thus, by using an interval $[0,T]$ that verifies the condition:
\begin{eqnarray}
|R^\gamma_{\lambda\mu\nu}| \ll T^{-2} \ll (\omega_{i}^{a})^{2}\;,
\end{eqnarray}
for any of the components of the Riemann tensor and of the vector,
it is possible to obtain (\ref{interval}) and prove that the mean value of the energy-momentum tensor 
is isotropic. Thus, if
oscillations are fast compared to the curvature scale, the
average energy-momentum tensor takes the perfect fluid form for any locally inertial observer.

\section{Conclusions}

In this work we have considered homogeneous non-abelian vector fields in an expanding universe with
arbitrary potential term. By means of a
generalized version of the virial theorem, we have shown that for bounded and rapid evolution compared to the rate
of expansion, the average energy-momentum tensor is isotropic for any kind of initial configuration of the fields. 
This result holds irrespective of the complicated dynamics of the coupled fields in non-abelian theories 
even 
in the presence of gauge-fixing terms, and it can be extended for any locally inertial observer in
arbitrary geometries provided the field evolution is sufficiently rapid.

These rapidly oscillating vector field models not only avoid the problem of
anisotropies at the classical level as shown in this work, but also could solve it 
even in the presence of quantum fluctuations. Indeed, they open up the possibility of using non-abelian fields in cosmology
beyond the simple triad configurations considered so far. 
Notice that light vector fields during inflation develop a classical homogeneous vacuum expectation 
value due to the sum of infrared modes. The quantum fluctuations of the fields in the triad will not
be correlated generally  and  will break   isotropy \cite{Peloso2}. However, in the cases considered
in this work, the high effective mass of vector perturbations compared to the rate of expansion
prevents the  long-wavelenght modes from being excited, thus avoiding the problem
of anistropy even at the quantum level.

\vspace{0.2cm}

{\bf Acknowledgements:}
We thank Marco Peloso and Jose Beltr\'an Jim\'enez for useful comments. This work has been supported by MICINN (Spain) project numbers FIS2011-23000, FPA2011-27853-01 and Consolider-Ingenio MULTIDARK CSD2009-00064.

\end{document}